Title: Ventricular torsional relation to ventricular fiber arrangement

Authors: Saeed Ranjbar [1,*], Ph.D, Tohid Emami Meybodi [2], MD, , Mahmood emami Meybodi [2], MD

**Research institute:**

1- Modarres Hospital, Institute of Cardiovascular Research, Shahid Beheshti University of Medical Science, Tehran, Iran.

2- Shahid sadoughi university of medical sciences, Yazd, Iran

***Corresponding Author:** Saeed Ranjbar

The full postal address of the corresponding author:

Modarres Hospital, Institute of Cardiovascular Research, Shahid Beheshti University of Medical Science, Tehran, Iran

E-mail: sranjbar@ipm.ir

Tel/FAX: +9821 22083106

Abstract:

Left ventricular torsion from helically oriented myofibers is a key parameter of cardiac performance. Physicians observing heart motion on echocardiograms, during cardiac catheterization, or in the operating room, are impressed by the twisting or rotary motion of the left ventricle during systole. Conceptually, the heart has been treated as a pressure chamber. The rotary or torsional deformation has been poorly understood by basic scientists and has lacked clinical relevance. The aim of this paper attempts to discuss about this question: Is ventricular twisting related to ventricular fiber arrangement? That is dependent to an assumed model of the left ventricular structure.

# Ventricular torsional relation to ventricular fiber arrangement

The twisting motion of the heart is believed to be secondary to the arrangement of the muscle fibers. Pettigrew [1] performed careful dissection of the heart of mammals and man, demonstrating 7 muscle layers. The three outer layers spiral with an increasing angle from the perpendicular, while the fourth layer is horizontal. The three inner layers spiral in the opposite direction, increasing toward the vertical. The layers are arranged in opposition so that 1 opposes 7, 2 opposes 6, and 3 opposes 5, with the fourth layer being a connecting layer. Pettigrew postulated that one triplet of muscles contracts during systole and the other stores energy that is utilized in diastole. In his view, the motion of the heart muscle is like that of a torsional pendulum. A reproduction of one of his anatomic dissections can be seen in Fig.1 [38].

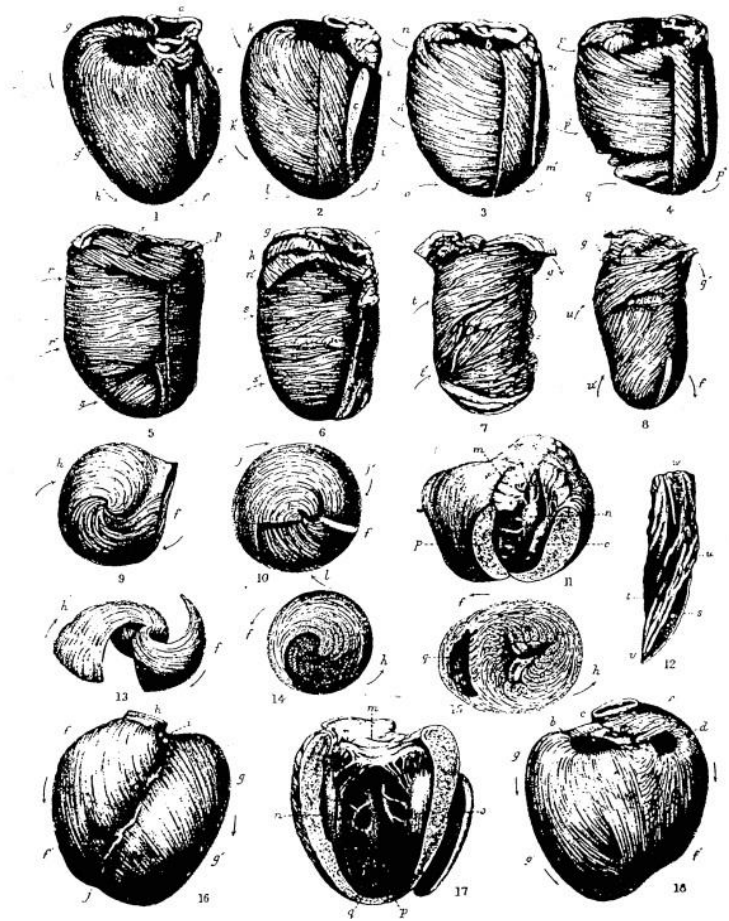

Figure 1: Anatomical dissections

Myocardial fiber orientation was examined by Streeter et al. [2] He reported that there was a well-ordered distribution varying from 60° (from the circumferential axis with positive being toward the base) on the endocardium to approximately -60° on the epicardium of the heart. He found that this fiber angle increased in systole by approximately 7° near the base and 19° near the apex relative to their counterparts in diastole, suggesting a torsion during contraction of die left ventricle.

More recently, Fernandez-Tran and Hurle [3] confirmed that there are three muscle fiber layers in the heart wall based on orientation. They described superficial (subpicardial), middle and deep (subendocardial), muscle layers which are similar for both the left and right ventricle with the exception of the middle layer which is found only in the left ventricle.

The opposing epicardial and endomyocardial muscle bundles might be important functionally. If potential energy is stored in these muscle fibers, the sudden release might promote filling during early diastole or the period of rapid ventricular filling. The proposed muscle bundles would also give a morphological correlate to the twisting or wringing motion of the left ventricle.

Since the discovery of the helical ventricular myocardial band by Francisco Torrent-Guasp >50 years ago [4] and the functional impact of the myocardial band [5, 6], many scientists have debated the validity of this concept. In any case, Torrent-Guasp's helical spiral concept may find an ideal connection with the spiralization of the outflow tracts and great arteries probably being related to the asymmetric intracardiac flow [7, 8] and/or to the spiral pattern detected at cellular/molecular level [9]. If the normal rightward spiralization represents the best hydrodynamic pattern, our first aim should be to try to restore the geometric pattern as close as possible to normality, avoiding, as it sometimes happens, the most abstruse surgical choice. In conclusion, the form and function of the heart are inevitably interdependent and we believe this to be true at each phylogenetic and ontogenetic stage. To treat heart diseases as best we can, we must try to understand the structure, function and deep mechanisms of the normal heart from its 'origin'. Once we have acquired as much knowledge as possible in this enormous field, we will then need to do the most difficult thing: mimic nature! [39]

Models of heart structure

Eight conceptual models of cardiac structure are summarized in Fig. 2 and Fig. 3. The variety of proposed structures is visually striking this is due, in a large part, to the different levels of cardiac structure described. Model 6 (Fig. 2, part 6) groups the myocardium into regional functional units analogous to skeletal muscles. Models 1—5 (Fig. 2, parts 1—5) are continuum concepts stressing to differing degrees the anisotropic interconnectivity inherent in cardiac structure. Models 1 and 2 (Fig. 2, parts 1 and 2) describe fiber orientation. Model 3 (Fig. 2, part 3) describes changes through layers from epicardium to endocardium. Models 4 and 5 (Fig. 2, parts 4 and 5) examine the myolaminar structure, with less emphasis on fibre orientation. Some descriptions of Model 7 (Fig. 2, part 7), the HVMB, have emphasized discrete structural bundles [10] while other reports have proposed the band within a continuum framework [11]. Model 8 describe a dynamic orientation contraction (through the cardiac cycle) of every individual myocardial fiber could be created by adding together the sequential steps of the multiple fragmented sectors of that fiber. This way we attempted to mechanically illustrate the global LV model Fig. 3. We serve to describe explicitly three models of heart that one has been provided by authors. [40]

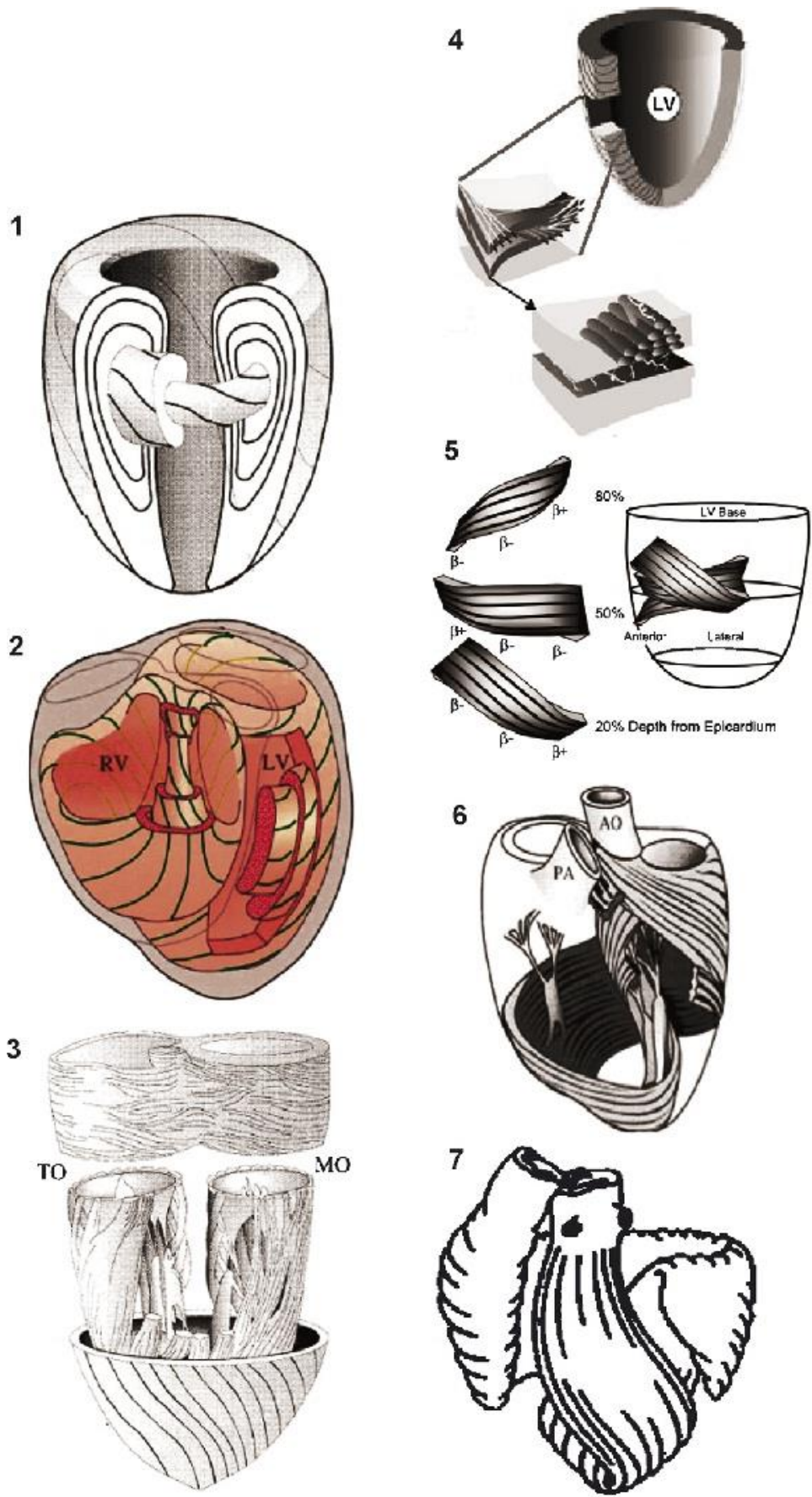

Figure 2: Mechanisms of heart in different models

Doughnut and pretzel models — Models 1 and 2

(Fig. 2, parts 1 and 2)

The fiber path models are based on the synthesis of many measurements of angles of fiber orientation through the ventricular wall and differ primarily in their scope (Model 1 is left ventricle (LV) only; Model 2 is LV and right ventricle (RV)). No argument is made for discrete traceable fiber paths — rather a recognizable general fiber trajectory [12]. These models reconstruct the long accepted helical pattern of fiber orientation, which has recently been confirmed by DT-MRI [13, 21-26] (Fig. 3). DT-MRI fiber tracing algorithms (Zhukov and Barr [14] and Kondratieva et al. [15], applied to canine data; Schmid et al. [13], applied to porcine data; Rohmer et al. [16], applied to human data) and automatic computational constructive visualization methods (Chen et al. [17,18]), used to extract meaningful visual information from histologically recorded three-dimensional fiber orientation datasets [19,20] (recorded from the rabbit heart), produce models remarkably similar to that proposed by Streeter (Fig. 2, part 1).

Peskin [27] has carried out an asymptotic analysis of Model 1 and has derived the fiber architecture of the heart from first principles.

These models form a conceptual model of cardiac structure from the one macro- and microscopically observed characteristic of principal fiber direction. As such, the nested doughnuts/pretzel surfaces are abstract concepts no discrete biological equivalents exist. It is possible to create the surfaces by dissection following the observed principal fiber direction, as demonstrated in early studies by Torrent-Guasp [12,28], but in so doing much information is lost. The dissection and histological methods used do not record data of local branching in directions other than the predominant fiber orientation, so by definition no detail of local tissue organization is modelled. Locally branching fibers are smoothed to a single orientation, which is again smoothed to a global ventricular fiber orientation. With reference to Grant's principles, these models represent the highest order schema of whole heart fiber orientation in isolation. It might be assumed that the fiber maps produced by algorithmic analysis and computer visualization may be more than a conceptual model, but the considerations above apply equally. The DT-MRI fiber-tracing algorithms track the principle fiber orientation from the primary eigenvector only [40]. The automatic computational constructive methods were applied to histological fiber orientation datasets which only record principle fiber orientation. Being limited to tracking the principle fiber orientation alone, these methods cannot reconstruct any detail of myolaminar structure, and as such reproduce idealized fiber-tracing dissections. Much evidence points to myolaminae as the central feature of the wall motion mechanism [29-33]. As such the doughnut and pretzel models represent geometric abstractions of cardiac structure. Their primary uses may be in (i) refining more histologically detailed models for the constraint of fiber orientation and (ii) in modeling the spread of the cardiac action potential, the conduction of which is significantly influenced by principal fiber direction [34].

Model 8 (Fig.3)

This model provides a dynamic orientation contraction (through the cardiac cycle) of every individual myocardial fiber could be created by adding together the sequential steps of the multiple fragmented sectors of that fiber. Our study shows that in normal cases myocardial fiber paths initiate from the posterior-basal region of the heart, continues through the LV free wall, reaches the septum, loops around the apex, ascends, and ends at the superior-anterior edge of LV [35, **41**].

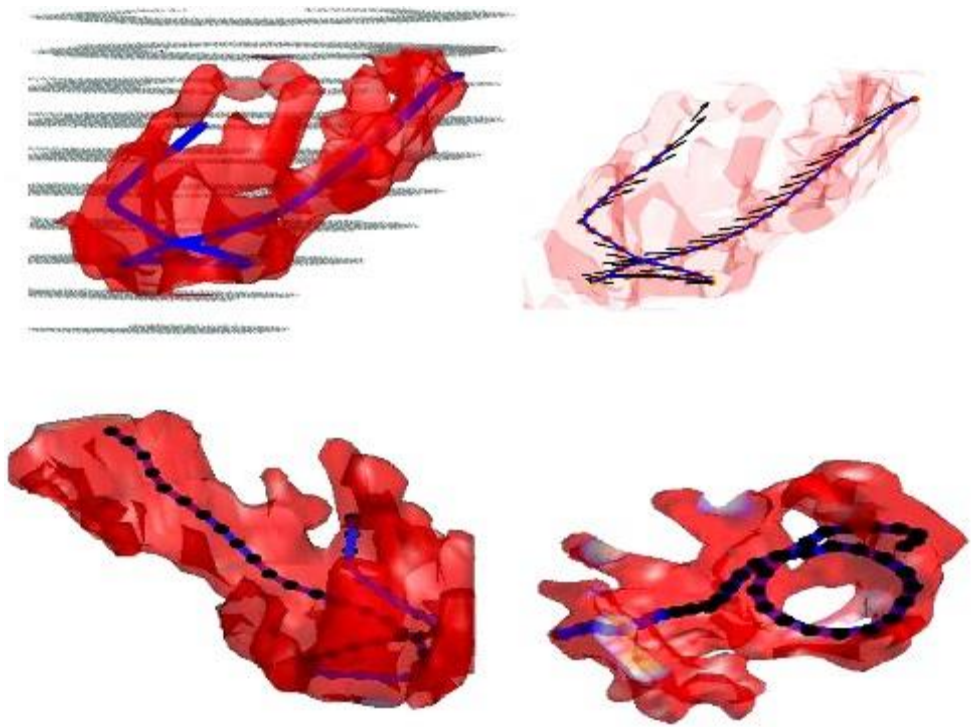

Figure 3: The rout of a myofiber in the left ventricle

# Philosophical approach for description of a complex structure

How can so many models co-exist? Some must be wrong or is it possible that different models are true representations of the heart when considered from different perspectives?

One 1965 review on cardiac structure stands out as a philosophical foundation upon which all future studies should be based. Robert P. Grant, M.D., in his Notes on the Muscular Architecture of the Left Ventricle [36], puts forward the following principles (along with an excellent description of cardiac structure):

(1) As evidenced from numerous dissections, a given fiber segment has branching connections with other fiber segments in several different directions —as such the myocardial structure is a syncytium-like arrangement [37].

(2) The cardiac structure problem is therefore a three dimensional network problem,

(3) Structure understood from study of the network depends on the level on which the structure is approached; whether an attempt is made to describe the predominant behavior in the entire ventricle(s) or at the other extreme, to describe the average branching from a specific location,

(4) Many models of heart structure can therefore be proposed depending on the conceptual approach,

(5) Statistical study of the branching must be added to geometry for an accurate picture of myocardial architecture,

(6) In consideration of such statistical problems one approach is to construct models for different degrees of generalization of the problem,

(7) No single model gives the whole story but together they provide a schema,

(8) The existence of a syncytium-like arrangement does not in itself dictate that no separate 'bundles' are present; however, due to the complex structure it is possible to construct by dissection bizarre arrangements which have no underlying anatomical reality.

(9) Any dissection of the myocardium may represent (a) a valid schema from an infinite set of valid schema, (b) a bizarre and meaningless pathway or (c) a grouping of fiber paths within the syncytium of such general shared fiber direction that it can be considered a physiological or anatomical bundle,

(10) A unique anatomical bundle is not the same as a unique physiological bundle — a connection between separate anatomical structures may produce one physiological structure,

(11) A statistical approach is required to demonstrate any independent anatomical entity.

When viewed from this perspective it is not surprising that many models exist, that these are not all mutually compatible and that argument continues. Further evidence for these principles is presented in a dissection study by Fox and Hutchins [37], where principle (1) is re-stated and emphasized: 'The only level of the network of cells that can be referred to accurately as a fiber is a single cell. The 'fiber' is often only one cell in length before it splits and branches'. When considering the structural debate it is logical to return to Grant's principles — an approach adopted in this article. It should be noted that although Grant suspected regional variations in fiber branching within the left ventricle, he was special whether these local prevalence would statistically warrant consideration as separate bundles. One feature of the myocardial structure problem Grant did not consider was that significant structural differences may exist between individuals of the same species. [40]

Conclusion:

Studies to date indicate that the left ventricle undergoes a torsional deformation, twisting in a counterclockwise direction during systole and then returning in a clockwise direction during diastole. LV myocardial models enable physicians to diagnose and follow-up many cardiac diseases and the torsional, deformation is likely related to the myocardial models. The opposing bands of muscle in the ventricular wall appear to be morphologically correlated with function.

Disclosure:

There is no disclosure.